\documentclass[12pt,preprint]{aastex}
\def\t0{\theta_{\circ}}

\def\be{\begin{equation}}
\def\en{\end{equation}}

\begin{document}

\title{Unraveling Brown Dwarf Origins}

\author{Ray Jayawardhana}
\affil{Department of Astronomy, University of Michigan, Ann Arbor, MI 48109, USA}

Brown dwarfs are not massive enough to be stars, nor are they planets in the traditional sense. They are often found in isolation but may also be companions to stars. They share some characteristics with stars and others with planets, and occupy the mass range in between: from about 75 to about 10 Jupiter masses. Although there is the potential for an identity crisis (1), brown dwarfs may hold clues to the formation of stars as well as planets. The identification of hundreds of brown dwarfs over the past 5 years in the solar neighborhood and in young star clusters (see the figure) has led to widespread interest in their origin. [For a discussion of brown dwarf chemistry, see (2).]

How can free-floating objects with inferred masses as low as 10 Jupiter masses form? One possibility is that brown dwarfs form as a result of the turbulent fragmentation and contraction of molecular cloud cores--that is, in the same way as stars. In 1976, Low and Lynden-Bell (3) calculated that the Jeans mass (the minimum mass for a cloud fragment to overcome pressure and collapse under gravity) could be as small as about 7 Jupiter masses, comparable to the masses of some planets detected around nearby Sun-like stars and below the threshold of 13 Jupiter masses for deuterium burning. This deuterium-burning limit is taken by some astronomers as the boundary between brown dwarfs and planets. Simulations suggest that fragmentation of a collapsing cloud might produce self-gravitating objects with initial masses as low as 1 Jupiter mass, provided that magnetic field tension effects are important in the cloud environment (4).

In another scenario proposed by Reipurth and Clarke (5), brown dwarfs are stellar embryos that are ejected from newborn multiple systems before they have accreted sufficient mass (about 75 Jupiter masses) to burn hydrogen. In this scenario, a stellar embryo competes with its siblings to accrete infalling matter; the one that grows slowest is the most likely to be ejected through dynamical interactions. In computer simulations, objects with very low mass are indeed often kicked out of nascent stellar systems (6). It has also been suggested that brown dwarfs form in gravitationally unstable regions of disks surrounding protostars or young binaries.

Studies of young brown dwarfs may help to distinguish among these different scenarios. For example, if the ejection hypothesis were true, one might expect disks around brown dwarfs to be pruned by close interactions within a multiple system (6). The disks may therefore be short-lived. It is also unlikely that many brown dwarfs would come in pairs, because such binaries are likely to be torn apart during ejection.

Several groups have recently searched for dusty disks around large numbers of young brown dwarfs in nearby star-forming regions and clusters (7-10). Such disks would absorb light from the central star or brown dwarf and re-emit at longer wavelengths, making objects with disks appear brighter in the infrared than those without. The results show that disks are rather common around brown dwarfs at an age of about 1 million years, as is the case for Sun-like stars. It has not yet been possible to measure the sizes of brown dwarf disks to determine whether they are truncated, but their lifetimes appear to be comparable to those of disks around stars.

Brown dwarfs also accrete material from surrounding disks in the same way as their stellar cousins. Using high-resolution optical spectrographs on the Keck and Magellan telescopes, astronomers have detected broad, asymmetric emission lines of hydrogen in the spectra of many very young brown dwarfs (11-13). These lines are the telltale sign of gas flowing from the inner edge of the disk onto the central object at velocities of 100 km/s or higher. The rate of mass accretion from the disk is lower in brown dwarfs than in solar-mass stars by a factor of 10 to 100. But in some cases, the material may continue to trickle in for up to 10 million years (14)--another piece of evidence for long-lived disks around these substellar objects.

A particularly intriguing case is the young brown dwarf LS-RCrA 1 in the R Corona Australis star-forming region. Its spectrum not only exhibits signs of ongoing accretion onto the brown dwarf, presumably from a disk, but also shows hints of mass outflow from it: It contains many forbidden emission lines (15, 16), which are usually associated with young stars in which a fraction of the inflowing material is ejected perpendicular to the disk. If confirmed through future observations, this finding would further strengthen the analogy between nascent brown dwarfs and their stellar counterparts.

The mounting evidence thus points to a similar infancy for Sun-like stars and brown dwarfs. Does this mean that the two kinds of objects are born in the same way? Many observers tend to think so (7-12), but it may be too early to rule out the ejection scenario for at least some brown dwarfs. Far-infrared observations with the Spitzer Space Telescope (launched in August 2003) and millimeter observations with ground-based radio telescopes may reveal the sizes and masses of brown dwarf disks, allowing us to determine whether most disks are truncated. Better statistics of the frequency of binary brown dwarfs could provide another observational test. Infrared studies of even younger ``proto-brown dwarfs,'' which are still embedded in a dusty womb, may also provide clues to their origin.

\newpage
{\bf References}

\noindent 1. G. Basri, Mercury 32, 27 (2003). \\
\noindent 2. K. Lodders, Science 303, 323 (2004).\\ 
\noindent 3. C. Low, D. Lynden-Bell, Mon. Not. R. Astron. Soc. 176, 367 
(1976).\\
\noindent 4. A. Boss, Astrophys. J. 551, L167 (2001). \\
\noindent 5. B. Reipurth, C. Clarke, Astron. J. 122, 432 (2001). \\
\noindent 6. M. R. Bate, I. A. Bonnell, V. Bromm, Mon. Not. R. Astron. Soc. 332, L65 (2002). \\
\noindent 7. A. A. Muench et al., Astrophys. J. 558, L51 (2001). \\
\noindent 8. A. Natta et al., Astron. Astrophys. 393, 597 (2002). \\
\noindent 9. R. Jayawardhana et al., Astron. J. 126, 1515 (2003). \\
\noindent 10. M. Liu et al., Astrophys. J. 585, 372 (2003). \\
\noindent 11. R. Jayawardhana, S. Mohanty, G. Basri, Astrophys. J. 578, L141 (2002). \\
\noindent 12. R. Jayawardhana, S. Mohanty, G. Basri, Astrophys. J. 592, 282 (2003). \\
\noindent 13. J. Muzerolle et al., Astrophys. J. 592, 266 (2003). \\
\noindent 14. S. Mohanty, R. Jayawardhana, D. Barrado y Navascu\'es, Astron. Astrophys. 380, 264 (2001). \\
\noindent 15. M. Fern\'andez, F. Comer\'on, Astron. Astrophys. 380, 264 (2001). \\
\noindent 16. D. Barrado y Navascu\'es, S. Mohanty, R. Jayawardhana, Astrophys. J., in press.\\

\end{document}